\documentclass[11pt]{article}
\usepackage{amsmath}
\usepackage{graphicx}
\usepackage{color}
\usepackage{orcidlink}
\usepackage{hyperref}
\hypersetup{colorlinks=true, linkcolor=blue, citecolor=blue, urlcolor=blue}
\usepackage{subcaption}
\RequirePackage[numbers,sort&compress]{natbib}
\begin{document}
\baselineskip=14pt

\begin{center}
\LARGE{Thermodynamics and shadow of Simpson-Visser black hole with phantom global monopoles}    
\end{center}

\vspace{0.3cm}

\begin{center}

{\bf Ahmad Al-Badawi}\orcidlink{0000-0002-3127-3453}\\
Department of Physics, Al-Hussein Bin Talal University, 71111,
Ma'an, Jordan. \\
e-mail: ahmadbadawi@ahu.edu.jo\\

\end{center}

\vspace{0.15cm}
\begin{abstract}
We investigate the thermodynamics and shadow of a non-rotating Simpson–Visser black hole with a phantom global monopole. The model is governed by three parameters: the coupling constant $\xi$, the energy scale of symmetry breaking $\eta$, and the bounce parameter $a$, which jointly influence horizon structure and observational signatures. Using specific heat and free-energy analysis, we show that small-horizon configurations are locally thermodynamically stable but never globally favored. Analytical solutions of null geodesics reveal that the photon sphere radius   depends on the bounce parameter $a$ and the energy scale of symmetry breaking $\eta$, while the critical impact parameter is still unaffected by $a$. Moreover,  the photon sphere radius and critical impact parameter, showing that increasing $\eta$ enlarges both quantities for an ordinary global monopole, while reducing them in the phantom case. Our results highlight how the bounce parameter and phantom global monopole significantly alter the black hole's physical and geometric properties.
\end{abstract}

{\bf keywords}: Simpson–Visser black hole;  Phantom global monopoles; Thermodynamic Properties; Shadow.

\vspace{0.3cm}

\section{Introduction} \label{sec1}

Traditional black holes (BHs) like the Schwarzschild and Reissner-Nordström types have two singularities: a coordinate singularity (the event horizon) and an essential singularity at $r=0$, where curvature becomes infinite. Regular BHs avoid the essential singularity, keeping curvature finite everywhere. They form when the strong energy condition is violated near the BH. There are two main approaches to constructing regular BHs: Changing the matter source (e.g., exotic matter distributions) and solving Einstein’s equations \cite{PN,EP,PN2,AB,ZR,AB2}. Second approach is to introduce quantum corrections to classical BH solutions \cite{LM,RG,AP,MB,NB1,SB}. The first regular BHs model was proposed by Bardeen \cite{BARDEEN}, later explained via field theory by Ayón-Beato-García \cite{ELOY}. Research has since expanded to include both non-rotating and rotating regular BHs \cite{VS1,VS2,VS3,VS4}.

The Simpson–Visser (SV) metric \cite{VS5}, a regularized spacetime model featuring a bounce parameter $a$, provides a framework for exploring non-singular BHs and wormholes while keeping key phenomenological features. This framework provides a unified description of regular BHs and wormholes by smoothly interpolating between the two via a length-scale parameter $a$, which regularizes the central singularity and may encapsulate quantum gravity effects in a minimally complex manner. A rotating version of this model has also been developed \cite{VS6}. The SV wormhole is traversable, much like the classic Morris-Thorne wormholes \cite{VS5}, functioning as a spacetime tunnel that connects distant regions of the Universe-or even separate universes-allowing the passage of massive objects. Meanwhile, their regular BH solution exhibits a "black bounce" behind the event horizon, resembling the concept of a black universe \cite{VS7}.

On the other hand, BHs with nontrivial topological defects, such as global monopoles (GMs), have garnered significant attention in   gravitational physics due to their intriguing implications for thermodynamics, geometry, and observational signatures. GM arise from spontaneous symmetry breaking in the early cosmos, where a global $SO(3)$ symmetry is disrupted, yielding different topological configurations \cite{isz10, isz11}. These imperfections alter spacetime geometry by contributing energy, causing solid angular deficits that affect geodesic motion and other physical processes \cite{isz12,isz13}. The influence of GMs on BH solutions has been extensively studied in General Relativity (GR) and modified gravity theories. Research spans various models, including static and spherically symmetric BHs with Kalb-Ramond fields  \cite{AA1}, Eddington-inspired Born-Infeld BHs  \cite{AA2}, and BHs in $f(R)$  \cite{AA3}, bumblebee  \cite{AA4}, and Gauss-Bonnet gravity  \cite{AA6}. These studies reveal rich thermodynamic behavior, phase transitions (e.g., van der Waals-like structures), and modified wave dynamics (e.g., greybody factors, QNMs).

In this work, we investigate the thermodynamics and shadow of a non-rotating SV BH imbued with a phantom GM. The SV metric when coupled with a GM the resulting spacetime exhibits modified gravitational properties, depending on whether the monopole is ordinary or phantom in nature. Our study is driven by the understanding that the interaction between SV spacetime and topological defects gives rise to rich physical phenomena beyond the predictions of standard GR. By introducing both a bounce parameter and GMs into the SV BH solution, we investigate how these modifications collectively shape key aspects of BH physics, particularly thermodynamics and shadow properties.

The paper is organized as follows: Section \ref{sec:2} presents SV BHs with PGMs, analyzing their metric structure,  event horizons, and scalar invariants. In Section \ref{sec:4}, we examine the BH's thermodynamics, deriving the Hawking temperature, entropy, Gibbs free energy, and  specific heat to assess thermal behavior as well as global and local  thermodynamic stability. Section \ref{sec:5} investigates the shadow properties of the BH. Finally, Section \ref{sec:6} summarizes our results, discusses their implications for BH physics and modified gravity, and identifies potential observational signatures that could distinguish SV BHs with phantom GMs (PGM) from classical solutions.

\section{ Non-rotating  Simpson–Visser BHs with PGM} \label{sec:2}

The concept of regular BHs was first proposed by Bardeen in his seminal work \cite{BARDEEN}. Since then, the topic has garnered significant attention in gravitational physics. A notable example of such a spacetime is the SV metric \cite{VS5}, which describes a non-rotating, static, and spherically symmetric regular BH. The metric is defined by
\begin{equation}\label{SV}
ds^2=-\left(1-\frac{2M}{\sqrt{r^2+a^2}}\right)\,dt^2
+\left(1-\frac{2M}{\sqrt{r^2+a^2}}\right)^{-1}\,dr^2 + (r^2+a^2)(d\theta^2+
sin^2\theta d\phi^2).
\end{equation}
Here, M is the ADM mass, and $a>0$ is the parameter having a dimension of length. Based on the value of the parameter $a$, this solution encompasses both regular BHs and wormholes.  The SV metric may represent three different types of
 spacetimes
\begin{itemize}
    \item For $a > 2M$, we have a two-way, traversable wormhole.
    \item For $a = 2M$, we have a one-way wormhole with a null throat.
    \item For $a <2M$, it results in a regular BH. In this situation, the singularity at $r=0$ is replaced by a bounce to another universe. In \cite{VS5}, black-bounce refers to the bounce that occurs through a space-like throat covered by an event horizon. 
\end{itemize}

Now let us provide a brief outline of a static, spherically symmetric PGM configuration that results from the spontaneous symmetry breaking of a triplet of phantom scalar fields with global $SO(3)$ symmetry. The action describing the formation of the PGM is given by \cite{SC1}:
\begin{equation}
    \mathcal{S}=\int\,\sqrt{-g}\,d^4x\,\left[\frac{R}{2}-\frac{\xi}{2}\,(\partial^{\mu}\psi^{i})\,(\partial_{\mu}\psi^{i})-\frac{\lambda}{4}\,(\psi^i\,\psi^i-\eta^2)^2 \right]\quad (i = 1,2,3).\label{aa3}
\end{equation}

Here, $\psi^i$ represents the triplet of scalar fields, $\eta$ being the energy scale of symmetry breaking, and $\lambda$ is a constant. The coupling constant $\xi$ in the kinetic term is set to $\xi=1$, which corresponds to an ordinary GM (OGM) arising from scalar fields with positive kinetic energy \cite{isz10}. When the coupling constant is $\xi=-1$, the kinetic energy of the scalar field becomes negative, leading to the formation of a PGM. The metric for this BH is given by \cite{SC1,SC2}:
\begin{eqnarray}
    ds^2&=&-A(r,\eta)\,dt^2+\frac{dr^2}{A(r,\eta)}+r^2\,(d\theta^2+\sin^2 \theta\,d\phi^2),\nonumber\\
    A(r,\eta)&=&1-8\,\pi\,\eta^2\,\xi-2\,M/r.\label{aa4}
\end{eqnarray}

We will now present a metric ansatz for static, spherically symmetric SV BH solution, featuring PGMs:
\begin{equation}
    ds^2=-f(r)\,dt^2+\frac{dr^2}{f(r)}+h(r)\,(d\theta^2+\sin^2 \theta\,d\phi^2),
     \label{aa5}
\end{equation}
where the metric functions $f(r)$  and $h(r)$ are defined by
\begin{equation}
f(r) = 1 - 8\,\pi\, \eta^2\, \xi-\frac{2M}{\sqrt{r^2+a^2}}, \hspace{1cm} h(r)=r^2+a^2.\label{aa6}
\end{equation}

To determine whether the above BH solution (\ref{aa5}) is singular or non-singular, one can compute the Ricci scalar, $\mathcal{R}=g^{\mu\nu}\,R_{\mu\nu}$ and the Kretschmann scalar,  $\mathcal{K}=R^{\mu\nu\lambda\sigma}\,R_{\mu\nu\lambda\sigma}$. For the chosen space-time (\ref{aa5}), we find these scalar quantities 
\begin{eqnarray}
\mathcal{R}&=&\frac{6\,a^2\,M}{\left(r^2+a^2\right)^{5/2}}-\frac{2\,a^2}{\left(r^2+a^2\right)^{2}}+\frac{16\pi(2a^2+r^2)\,\xi^2\,\eta^2}{\left(r^2+a^2\right)^{2}},\label{scalar1}\\
\mathcal{K}&=&\frac{4}{\left(r^2+a^2\right)^{5}}\left(\sqrt{a^2+r^2}[8\,a^2M(r^2-a^2)]+3\,a^4(a^2+r^2)+3\,M^2(3\,a^4-4a^2r^2+4r^4) \right) \nonumber\\ &+& \frac{64\,\pi\,\xi\,\eta^2}{\left(r^2+a^2\right)^{9/2}}\left( a^2r^2(\sqrt{a^2+r^2}-2M)  +2r^4(M+2\pi\,\xi\eta^2\sqrt{a^2+r^2})+a^2(4M+2(4\pi\,\xi\eta^2-1)\sqrt{a^2+r^2})  \right) .\quad\label{scalar2}
\end{eqnarray}
From the above expressions, we observe that the scalar curvatures exhibits the following feature:
\begin{eqnarray}
    &&\lim_{r \to 0} \mathcal{R}=\frac{6M}{a^3}+\frac{32\,\pi\, \eta^2\, \xi-2}{a^2}\quad,\quad \lim_{r \to \infty} \mathcal{R}=0\nonumber\\
    &&\lim_{r \to 0} \mathcal{K}=\frac{4\left(    a^2(3+32\,\pi\, \eta^2\, \xi(4\,\pi\, \eta^2\, \xi-1))+M(9M+8a^2(8\,\pi\, \eta^2\, \xi-1))\right)}{a^6}\quad,\quad \lim_{r \to \infty} \mathcal{K}=0.\label{scalar2a}
\end{eqnarray}
\begin{figure}[ht!]
    \centering
 \includegraphics[width=0.6\linewidth]{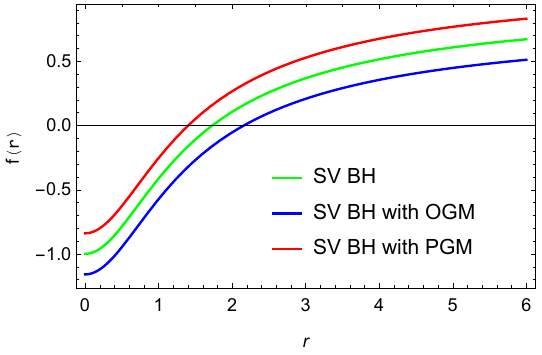}
    \caption{ A comparison of the metric function $f(r)$ for different BHs as a function of $r$. Here, $a/M=1$, $\eta=0.4$.}
    \label{figa1}
\end{figure}

The above  observations confirm that the SV BH with PGM metric (\ref{aa5}) under consideration 
represents a regular BH solution in GR.

The lapse function Eq. (\ref{aa6})  plays a crucial role in determining key observables associated with the underlying spacetime. In Figure \ref{figa1}, we present a comparative analysis of the metric function for different BH spacetimes, examining both OGM and PGMs as functions of the radial coordinate 
$r$. The plot reveals that, compared to the SV BH, the event horizon shifts outward in the case of OGM but inward for PGM.

\section{ Thermodynamics of Simpson-Visser BH with GMs  }\label{sec:4}

In this section, we examine the thermodynamic properties of the selected SV BH solution in the presence of GMs, encompassing both OGM and PGM types. Our investigation emphasizes the role of the bounce parameter $a$ and the GM structure determined by the symmetry-breaking energy scale in modifying key thermodynamic quantities. The introduction of GMs not only modifies the horizon structure but also significantly affects the thermodynamic stability of the BH. We conduct a detailed comparative analysis of these effects, distinguishing between OGM and PGM contributions, and assess their respective impacts on the evolution of thermodynamic behavior.

{In the framework of BH thermodynamics, the mass $M$ is interpreted as the enthalpy of the system. To derive other thermodynamic quantities, it is necessary to express the mass as a function of the entropy, which is itself proportional to the horizon area. The first step is to obtain the relationship between the mass M and the horizon radius $r_h$. This is achieved by solving the horizon condition $f(r_h)=0$, where $f(r)$ is the metric function defined in Eq. (\ref{aa6}), for the mass parameter. Solving this equation yields}
\begin{equation}
    M=\frac{1}{2}\sqrt{a^2+r_h^2}\left(1-8\,\pi\,\xi\,\eta^2  \right).\label{Mass1}
\end{equation}
In the presence of an OGM, the BH mass is positive if the  restriction 
$8\,\pi\,\eta^2\,<1$ is met. In the presence of a PGM, the BH mass is always positive. {Therefore, in the subsequent thermodynamic analysis, we will restrict our study to the parameter space where $M >0$ and a physical event horizon exists.}

Next, it is well-known that the Hawking temperature of a static spherically symmetric BH spacetime is given by $T=\kappa /2\pi$ \cite{SC3}, in which $\kappa=f'(r_{h})/2$ is the surface gravity of the BH on its event horizon. Therefore,
\begin{equation}
    T=\frac{1}{4\,\pi}\,\left(\frac{1-8\,\pi\xi\,\eta^2}{a^2+r_h^2}\right)r_h.\label{Temperature}
\end{equation}
In the presence of an OGM, the Hawking temperature  is positive for $8\,\pi\,\eta^2\,<1$. In the presence of a PGM, the Hawking temperature is always positive.

In the limit $\eta=0$, corresponding to the absence of GMs, the Hawking temperature from Eq. (\ref{Temperature}) reduces to the Hawking temperature of SV BH as \cite{VS5}
\begin{equation}
    T=\frac{1}{4\,\pi}\,\left(\frac{r_h}{a^2+r_h^2}\right).\label{Temperature2}
\end{equation}
Moreover,  in the limit $a=0$  (absence of the bounce parameter) and $\eta=0$ (absence of the GM), we recover the Hawking temperature for the standard Schwarzschild BH given by $T=\frac{1}{4\,\pi\,r_h}$, where $r_h=2\,M$. 

In Equation (\ref{Temperature}), it is evident that the PGM case ($\xi = -1$) has a much greater impact than OGM case ($\xi = 1$), which is evident from the following equation as,
\begin{equation}
\frac{T (\text{PGM})}{T(\text{OGM})}=\frac{1+8\,\pi\,\eta^2}{1-8\,\pi\,\eta^2}>1.\label{temperature3}    
\end{equation}

\begin{figure}[ht!]
\begin{center}
\subfloat[$\eta=0.4,\xi=1$]{\centering{}\includegraphics[scale=0.75]{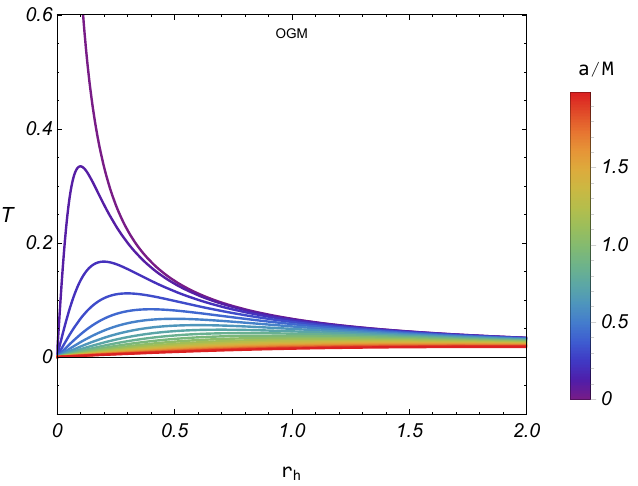}}\quad
\subfloat[$\eta=0.4,\xi=-1$]{\centering{}\includegraphics[scale=0.75]{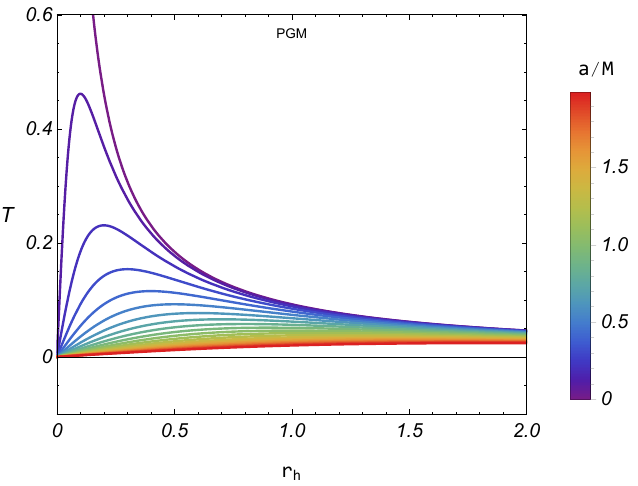}}
\end{center}
\caption{The behavior of the Hawking temperature for the SV BH with PGMs, highlighting the influence of the bounce parameter $a$.}\label{figa2}
\end{figure}
\begin{figure}[ht!]
\begin{center}
\subfloat[$a/M=1,\xi=1$]{\centering{}\includegraphics[scale=0.75]{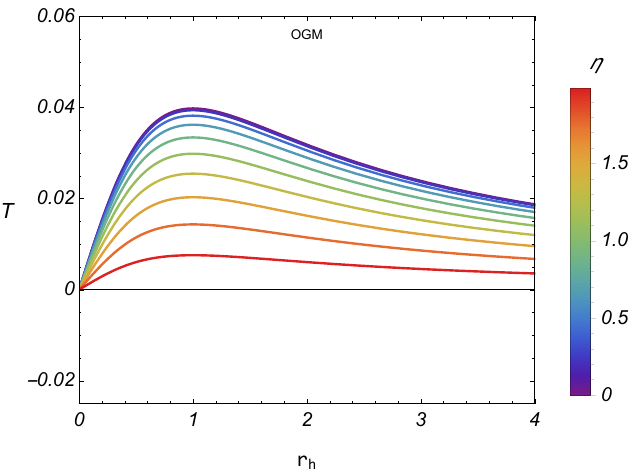}}\quad
\subfloat[$a/M=1,\xi=-1$]{\centering{}\includegraphics[scale=0.75]{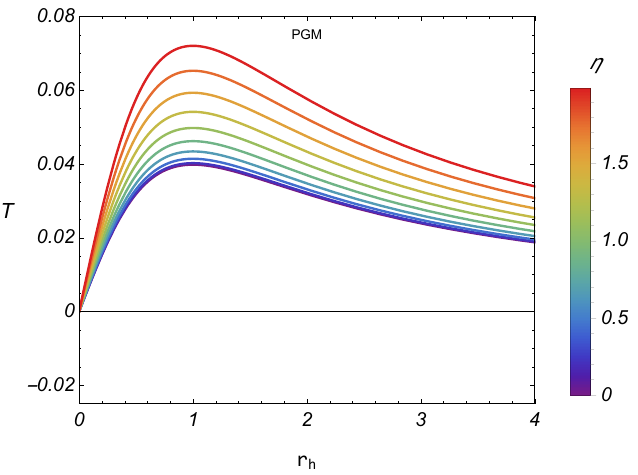}}
\end{center}
\caption{The behavior of the Hawking temperature for the SV BH with PGMs, highlighting the influence of the energy scale of symmetry breaking  parameter $\eta$.}\label{figa3}
\end{figure}

Figures \ref{figa2} and \ref{figa3} illustrate the behavior of the Hawking temperature as a function of the event horizon radius $r_h$, for varying values of the bounce parameter $a$ and the symmetry-breaking energy scale $\eta$ respectively. The figures show that as $r_h$ increases from its minimum value at $T = 0$, the temperature initially rises, reaches a maximum, and then falls at bigger horizon radii.  This local maximum causes a divergence in heat capacity, indicating a shift from thermodynamically stable to unstable BH branch. Moreover, the figures demonstrate that the PGM case has a much greater impact than OGM case.

 Next, let us  calculate the Bekenstein entropy using the four fundamental laws of BH thermodynamics: \begin{equation}
S=\int \frac{1}{T } \frac{\partial {M}}{\partial r_{h}}dr_{h}=\pi\left(r_h\sqrt{a^2+r_h^2}+a^2\,\text{log}\left[r_h+\sqrt{a^2+r_h^2}\right]   \right).
\label{entr2}
\end{equation}  

The expression in Eq. (\ref{entr2}) shows that the BH entropy is independent of the GM parameter $\eta$, but it does vary with the bounce parameter $a$ for a fixed event horizon radius $r_{h}$. In the case $a = 0$, which corresponds to the absence of bounce parameter in the selected BH solution, Eq. (\ref{entr2}) reduces to the conventional Bekenstein-Hawking entropy formula, $S = \pi\,r_h^2$. Moreover, the expression in Eq. (\ref{entr2})  indicates that the BH's entropy $S$ increases monotonically with the event horizon.

To assess global thermodynamic stability, we analyze the Gibbs free energy, which quantifies the maximum reversible work extractable from a system. Thermodynamically, it is defined as 
$G=M-T\,S$. By substituting equations (\ref{Mass1}), (\ref{Temperature}), and (\ref{entr2}), we derive the following expression for the Gibbs free energy:

\begin{equation}
    G=\frac{(8\,\pi\,\xi\,\eta^2\,-1)\left( a^2\,r_h \text{log}\left[r_h+\sqrt{a^2+r_h^2}\right]-(2a^2+r_h^2)\sqrt{a^2+r_h^2}  \right)}{4(a^2+r_h^2)}.\label{gibs1}
\end{equation}

In the limit $\eta=0$, corresponding to the absence of GMs, Eq. (\ref{gibs1}) reduces to  the Gibbs free energy of SV BH
\begin{equation}
    G=\frac{ (2a^2+r_h^2)\sqrt{a^2+r_h^2}-a^2\,r_h \text{log}\left[r_h+\sqrt{a^2+r_h^2}\right] }{4(a^2+r_h^2)}.\label{gibs2}
\end{equation}

Comparing Eqs. (\ref{gibs1}) and  (\ref{gibs2}), we observe that both the  bounce parameter and the symmetry-breaking energy scale simultaneously modify the Gibbs free energy compared to the standard Schwarzschild BH result. Notably, the impact of a PGM case ($\xi = -1$) on the Gibbs free energy in the considered BH is comparatively more than that of an OGM case ($\xi = 1$), which is clear from the following equation using (\ref{gibs1}) as,
\begin{equation}
\frac{G(\text{PGM})}{G(\text{OGM})}=\frac{1+8\,\pi\,\eta^2}{1-8\,\pi\,\eta^2}=\frac{T_H (\text{PGM})}{T_H(\text{OGM})}>1.\label{gibs4}    
\end{equation}

\begin{figure}[ht!]
\begin{center}
\includegraphics[scale=1]{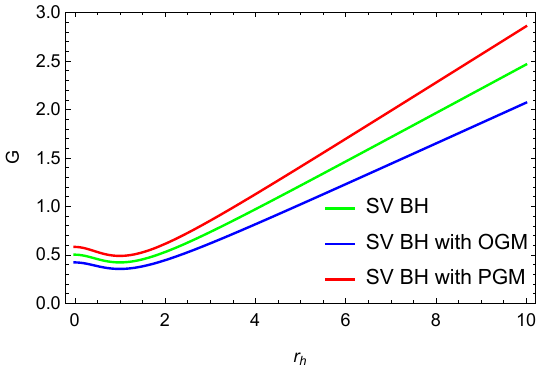}\quad
\end{center}
\caption{A comparison of the Gibbs free energy $G$ for different BHs as a function of $r$. Here, $a/M=1$, $\eta=0.4$.}\label{figGib}
\end{figure}

Global stability and phase transitions depend on the Gibbs free energy \cite{SC4,SC5}. The system is globally unstable when the free energy is positive and stable when it is negative. Figure \ref{figGib} compares the Gibbs free energy $G$ of different BH configurations (SV, SV with OGM, and SV with PGM) as a function of the radial coordinate $r$. The results demonstrate that in all scenarios, the Gibbs free energy remains positive, indicating that the SV BH, whether with or without GMs, is globally unstable.

Next, we aim to analyze the BH's local stability by examining its specific heat ($C$). This investigation will reveal whether the system undergoes phase transitions, as indicated by the behavior of its specific heat. A positive specific heat signifies stability during phase transitions, while a negative value suggests instability. The specific heat is calculated using the following standard relation: $C=\frac{dM}{dT}$  namely:  
\begin{equation}
   C=\frac{4\,\pi\,(a^2+r_h^2)^2}{(a^2-r_h^2)\left(1-8\,\pi\,\xi\,\eta^2\,r_h^2\right)}.\label{heat}
\end{equation}

From the expression given in Eq. (\ref{heat}), it is evident that the specific heat capacity is influenced by the bounce parameter $a$ and the symmetry-breaking energy-scale $\eta$ for a fixed event horizon radius. 

In the limit $\eta=0$, corresponds to the absence of the GM, the specific heat from Eq. (\ref{heat}) reduces to the specific heat of SV BH:
\begin{equation}
  C=\frac{4\,\pi\,(a^2+r_h^2)^2}{a^2-r_h^2}.\label{heat2}
\end{equation}
Moreover, in the limit $\xi=0=\eta$ ( absence of the bounce parameter and GM), the specific heat from Eq. (\ref{heat}) reduces to $C=-2\,\pi\,r_h^2$.

By comparing Eqs.~(\ref{heat}) and (\ref{heat2}), it is obvious that the bounce parameter $a$ and the symmetry-breaking energy scale $\eta$ alone or together affect the specific heat of the selected BH solution, resulting in variations from the typical Schwarzschild solution. In Equation \ref{heat}, it is evident that the OGM case ($\xi = 1$) has a much greater impact than PGM case ($\xi = -1$), which is evident from the following equation as,
\begin{equation}
\frac{C (\text{PGM})}{C(\text{OGM})}=\frac{1-8\,\pi\,\eta^2}{1+8\,\pi\,\eta^2}<1.\label{temperature4}    
\end{equation}

\begin{figure}[ht!]
\begin{center}
\subfloat[$\eta=0.4,\xi=1$]{\centering{}\includegraphics[scale=0.75]{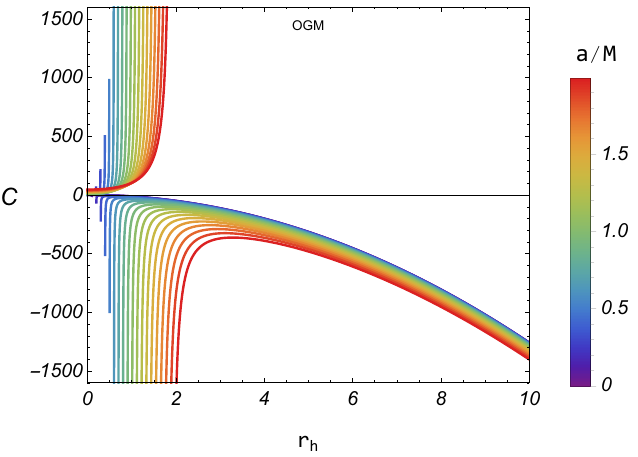}}\quad
\subfloat[$\eta=0.4,\xi=-1$]{\centering{}\includegraphics[scale=0.75]{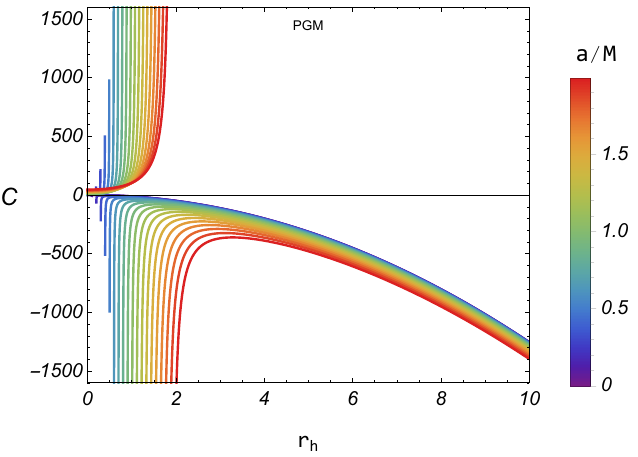}}
\end{center}
\caption{The behavior of the specific heat for the SV BH with PGMs, highlighting the influence of the bounce parameter $a$.}\label{figa2c}
\end{figure}
\begin{figure}[ht!]
\begin{center}
\subfloat[$a/M=1,\xi=1$]{\centering{}\includegraphics[scale=0.75]{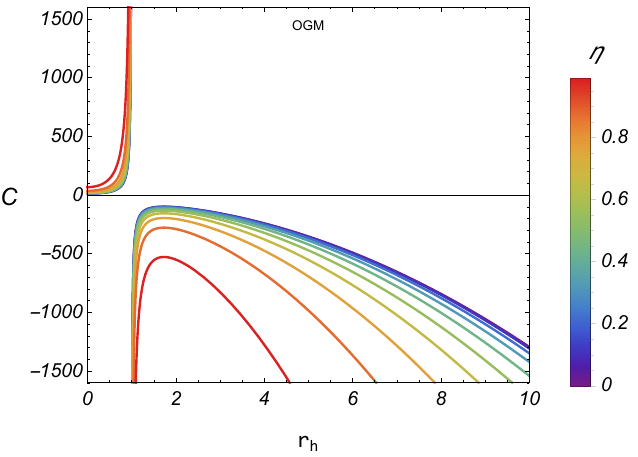}}\quad
\subfloat[$a/M=1,\xi=-1$]{\centering{}\includegraphics[scale=0.75]{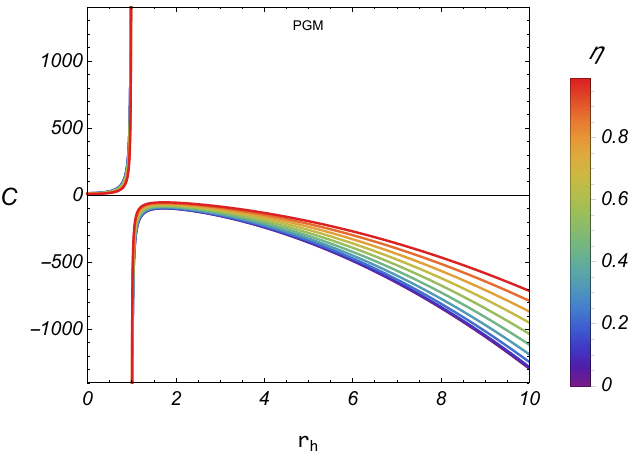}}
\end{center}
\caption{The behavior of the specific heat for the SV BH with PGMs, highlighting the influence of the energy scale of symmetry breaking  parameter $\eta$.}\label{figa3c}
\end{figure}

The sign of 
$C$
  is crucial for evaluating the thermodynamic stability of the BH. If $C>0$, the BH remains thermodynamically stable, whereas 
$C<0$ indicates an unstable solution. Therefore, the specific heat capacity is plotted as a function of the event horizon radius $r_h$ for different values of the bounce parameter 
$a$ and the energy scale parameter 
$\eta$ in Figures \ref{figa2c} and \ref{figa3c}. 

Figure \ref{figa2c} illustrates the dependence of the specific heat on the bounce parameter 
$a$ for OGM (left panel) and PGM (right panel). The plot reveals a single divergence point, signaling a phase transition in the thermodynamic behavior of the SV BH with GMs. The position of this divergence is sensitive to 
$a$: for smaller values of $a$, it occurs at smaller 
$r_h$, indicating a weaker influence of the bounce parameter.

Figure \ref{figa3c} displays the variation of specific heat with the energy scale parameter 
$\eta$ for OGM (left panel) and PGM (right panel). Similarly, a single divergence point appears, but the effect of 
$\eta$ is asymmetric compared to that of $a$.

\section{Shadows of Simpson-Visser BH with GM\MakeLowercase{s} } \label{sec:5}

{The investigation of BH shadows provides a crucial observational link between theory and experiment in strong-field gravity. The shadow, formed by photons that are captured or strongly lensed near the event horizon, carries detailed information about the underlying spacetime geometry and the parameters defining the BH. This investigation of the black hole shadow is motivated by its unique role as a direct probe of strong-field gravity and the fundamental parameters of BHs \cite{1804.05812, 1906.00873}. The landmark imaging of the M87* and Sgr A* shadows by the Event Horizon Telescope has transformed these theoretical constructs into observational tools, enabling unprecedented tests of general relativity and alternative gravity theories  \cite{1906.11238, 2311.08680}. Recent theoretical work continues to explore shadow properties in various modified gravity and matter environments  \cite{2501.14516, 2507.17280, 2506.19581}. Building upon this foundation, we here study the shadow cast by SV BH with PGM model to determine how PGM influences its apparent size and shape, thereby providing new observational signatures to distinguish between different theoretical scenarios. }\\To investigate the shadow cast, we must develop the equations of motion for photons (null geodesics) in this spacetime. To get the geodesic equations, we can use the Lagrangian density function  provided by \cite{sec3is01,sec3is02x,sec3is04,sec3is07}
\begin{equation}
2\,\mathcal{L}=-f(r)\dot{t}^2+\frac{\dot{r}^2}{f(r)}+h(r)\dot{\phi}^2.
\end{equation}
With the help of definition $p_q=\frac{\partial \,\mathcal{L}}{\partial \dot{q}}$, we obtain
\begin{eqnarray}\nonumber
p_t&=&\frac{\partial \,\mathcal{L}}{\partial \dot{t}}=-f(r)\dot{t}, \\\nonumber
p_r&=&\frac{\partial \,\mathcal{L}}{\partial \dot{r}}=\frac{\dot{r}}{f(r)}, \\
p_\phi&=&\frac{\partial \,\mathcal{L}}{\partial \dot{\phi}}=h(r)\dot{\phi}.
\end{eqnarray}
Here, the dot is differentiation with respect to an affine parameter $\tau$. Thus, in terms of energy $ \mathcal{E}$ and angular momentum $L$, we get two very important differential equations given by
\begin{equation}
\frac{d t}{d \tau}=\frac{\mathcal{E}}{f(r)} \quad \text{and} \quad \frac{d \phi}{d \tau}=\frac{L}{h(r)}.\label{Conserved}
\end{equation}
The equation for the null geodesics is
\begin{equation}
\left(\frac{d r}{d \tau}\right)^{2} \equiv \dot{r}^{2}
=\mathcal{E}^{2}-V(r),
\end{equation}
where $V(r)$ is the effective potential for photon and is  given 
\begin{equation}
V(r)=\frac{L^{2} f(r)}{h(r)}.
\end{equation}
The effective potential given above determines the motion of any particle in the underlying spacetime. 

For circular photon orbits of radius ($r_{ph}$), we must have
\begin{equation}
\frac{d V}{d r}=0\Rightarrow \frac{f^{\prime}(r_{ph})}{f(r_{ph})}=\frac{h^{\prime}(r_{ph})}{h(r_{ph})}.
\end{equation}
By inserting the metric functions in the previous equation, we get the equation for the photon sphere as 
\begin{equation}
    3\,M+(8\,\pi\,\eta^2\,\xi-1)\sqrt{a^2+r_{ph}^2}=0.
\end{equation}
The analytical solution of the photon sphere is 
\begin{equation}
    r_{ph}=\sqrt{\frac{9M^2}{\left(  1-8\,\pi\,\eta^2\,\xi\right)^2}-a^2}.\label{photons1}
\end{equation}

From expression given in Eq. (\ref{photons1}), it becomes evident that the radius of photon sphere $r_{ph}$ is influenced by the bounce parameter       $a$, the energy-scale of the symmetry   breaking $\eta$, and the BH mass $M$. In the limit $\eta=0$, that is without GMs, Eq. (\ref{photons1}) reduces to the photon sphere radius of SV BH, namely
\begin{equation}
   r_{ph}=\sqrt{9M^2-a^2}.\label{photons2} 
\end{equation}

\begin{figure}[ht!]
    \centering
    \includegraphics[width=0.45\linewidth]{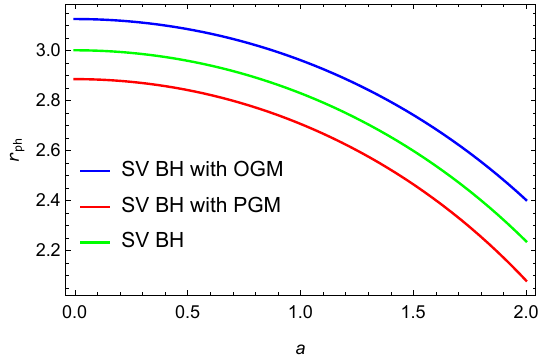}\quad\quad
    \includegraphics[width=0.45\linewidth]{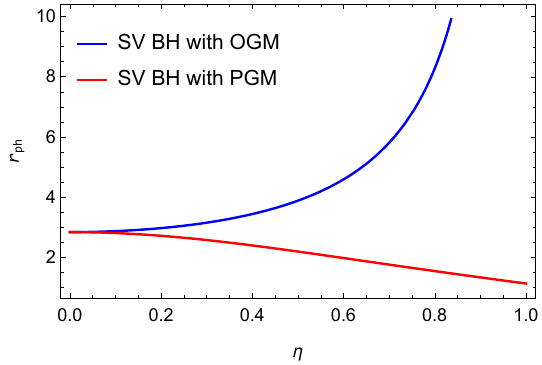}
    \caption{Plot of  the photon sphere $ r_{ph}$ versus  the  parameter $a$ (left) and versus the parameter $\eta$ (right).}
    \label{figph1}
\end{figure}

To demonstrate how the bounce parameter 
$a$ and the symmetry-breaking energy scale 
$\eta$ affect the photon sphere radius, we tabulate numerical values of $r_{ph}$ (Table \ref{table1a}) and present the results in Figure \ref{figph1}. The left panel of the figure shows the dependence of the photon sphere radius on the parameter 
$a$ for different SV BHs, while the right panel focuses on SV BHs with GMs. Our findings indicate that increasing 
$a$ consistently decreases the photon sphere radius across all cases. In contrast, the influence of 
$\eta$ is asymmetric: it enlarges the photon sphere radius in OGM but diminishes it in PGM.
\begin{center}
\begin{tabular}{|c|cc|cc|cc|cc|cc|}
 \hline \multicolumn{11}{|c|}{ Photon sphere radius ($r_{ph}$) }
 \\ \hline
  &\multicolumn{2}{|c|}{ $\eta=0.1$}&\multicolumn{2}{|c|}{ $\eta=0.2$}&\multicolumn{2}{|c|}{ $\eta=0.3$}&\multicolumn{2}{|c|}{ $\eta=0.4$}&\multicolumn{2}{|c|}{ $\eta=0.5$} \\ \hline 
$a/M $ & OGMs & PGMs & OGMs & PGMs & OGMs & PGMs & OGMs & PGMs & OGMs & PGMs
\\ \hline
$0.2$ & $3.0237$ & $2.96356$ & $3.11859$ & $2.87767$ & $3.29063$ & $2.74502$
& $3.56582$ & $2.57846$ & $3.995$ & $2.39165$ \\ 
$0.6$ & $2.97031$ & $2.90907$ & $3.06686$ & $2.82153$ & $3.24164$ & $2.6861$
& $3.52067$ & $2.51564$ & $3.95474$ & $2.32379$ \\ 
$1$ & $2.86055$ & $2.7969$ & $2.96068$ & $2.70574$ & $3.14138$ & $2.5642$ & $%
3.42857$ & $2.38505$ & $3.87298$ & $2.18174$ \\ 
$1.4$ & $2.68751$ & $2.61967$ & $2.79385$ & $2.5221$ & $2.98467$ & $2.36962$
& $3.28559$ & $2.1745$ & $3.747$ & $1.94936$ \\ 
$1.8$ & $2.43777$ & $2.36277$ & $2.55453$ & $2.25411$ & $2.76193$ & $2.0821$
& $3.08466$ & $1.857$ & $3.57211$ & $1.58745$ \\
 \hline
\end{tabular}
\captionof{table}{The photon sphere has  been tabulated numerically for  SV BH with GMs.} \label{table1a}
\end{center}

The corresponding impact parameter is
\begin{equation}
b_p=\frac{L}{\mathcal{E}}=\sqrt{\frac{h(r_p)}{f(r_p)}}\Rightarrow b_p= 3\,M\sqrt{\frac{3}{\left(  1-8\,\pi\,\eta^2\,\xi\right)^3}}, \label{photons10}
\end{equation}
{\begin{equation}
R_{\mathrm{s}}
=b_{\mathrm{p}}\sqrt{f(r_{o})}
={\frac{3\sqrt{3}\,M}{ 1-8\,\pi\,\eta^2\,\xi}}.
\end{equation}
which is the same value as that for  a BH solution with PGMs. Thus, we observe that while the photon radius depends on the bounce parameter 
$a$, the critical impact parameter remains independent of 
$a$. For a distant observer, the shadow radius coincides with the critical impact parameter, which implies that the shadow size itself is unaffected by variations in 
$a$.

To have a better understanding of the impact of the symmetry-breaking energy scale 
$\eta$ on the shadow radius, we tabulate numerical values of $R_{s}$ (Table \ref{table2a}) and present the results in Figure \ref{figsh01}. The figure reveals that increasing 
$\eta$ enlarges the shadow radius in the OGM scenario, while reducing it in the PGM case.
\begin{center}
\begin{tabular}{|c|c|c|c|c|c|}
 \hline \multicolumn{6}{|c|}{Shadow radius ($R_s$) }
  \\ \hline 
$\eta $ & $0.1$ & $0.2$ & $0.3$ & $0.4$ & $0.5$ 
\\ \hline
OGMs & $5.27508$ & $5.52427$ & $5.98577$ & $6.74937$ & $8.$  \\ 
PGMs& $5.11917$ & $4.89928$ & $4.56607$ & $4.15906$ & $3.71806$ \\ 

 \hline
\end{tabular}
\captionof{table}{The shadow radius  has  been tabulated numerically for  SV BH with GMs.} \label{table2a}
\end{center}

\begin{figure}[ht!]
    \centering
    \includegraphics[width=0.5\linewidth]{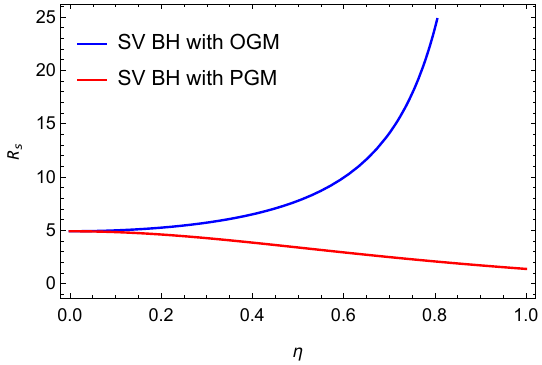}
    \caption{Plot of  the shadow radius $ R_{s}$ versus  the  parameter $\eta$ for SV BH with GMs. }
    \label{figsh01}
\end{figure} 
Let us introduce the celestial coordinates, $X$ and $Y$,  to represent the
actual shadow of the BH as observed from the observer's frame. {
For spacetimes that are not asymptotically flat the celestial coordinates must be defined with respect to a local observer at a large but finite distance 
$r_{\mathrm{o}}$. In this case, the coordinates are obtained by projecting the photon’s four-momentum onto the observer’s orthonormal tetrad \cite{new1,new3,new2}. %For null geodesics with conserved energy $\mathcal{E}$ and angular momentum $L$, the impact parameters are defined as $\zeta=L/\mathcal{E}$ and $\kappa=K/\mathcal{E}^2$, where $K$ is the Carter constant. The resulting expressions, 
{ The observer’s four-velocity is
\begin{equation}
u^\mu=\left(\frac{1}{\sqrt{f(r_o)}},0,0,0\right),
\end{equation}
For the metric ansatz given in Eq. (\ref{aa5}) we define celestial coordinates via a local orthonormal tetrad $e_{\hat{\mu}}$ for a static observer at $(r_o, \theta_o)$. The basis vectors are defined as:
and a convenient orthonormal tetrad at \((r_o,\theta_o)\) is \begin{equation*}
 e_{(t)}^{\ \mu} = \left(\frac{1}{\sqrt{f(r_o)}},0,0,0\right), \hspace{1 cm} e_{(r)}^{\ \mu} = \left(0,\sqrt{f(r_o)},0,0\right),
\end{equation*}
\begin{equation}
    e_{(\theta)}^{\ \mu} = \left(0,0,\frac{1}{r_o},0\right), \hspace{1 cm}
e_{(\phi)}^{\ \mu} = \left(0,0,0,\frac{1}{r_o\sin\theta_o}\right).  
\end{equation}

For a photon with conserved energy \(\mathcal{E}\), angular momentum \(L\), and Carter constant \(K\), we introduce the impact parameters \(\zeta=L/\mathcal{E}\) and \(\kappa=K/\mathcal{E}^2\). Projecting the photon four-momentum onto the tetrad yields
\begin{equation}
p^{(t)}=\frac{\mathcal{E}}{\sqrt{f(r_o)}}, \qquad
p^{(\phi)}=\frac{L}{r_o\sin\theta_o}, \qquad
p^{(\theta)}=\frac{\sqrt{\Theta(\theta_o)}}{r_o},
\end{equation}
where
\begin{equation}
\Theta(\theta)=\kappa+a^2\cos^2\theta-\zeta^2\cot^2\theta.
\end{equation}
The celestial coordinates \((X,Y)\) on the observer’s sky are defined as
\begin{equation}
X=-r_o\frac{p^{(\phi)}}{p^{(t)}}, \qquad
Y=r_o\frac{p^{(\theta)}}{p^{(t)}},
\end{equation}
which immediately lead to
\begin{equation}
X=-\frac{\zeta}{\sqrt{f(r_o)}}\,\csc\theta_o,
\qquad
Y=\pm\sqrt{\frac{\kappa+a^2\cos^2\theta_o-\zeta^2\cot^2\theta_o}{f(r_o)}}.
\end{equation}
These expressions reduce to the standard asymptotically flat results when \(f(r_o)\to1\) and explicitly demonstrate the observer dependence of the shadow in non–asymptotically flat spacetimes.}

We now analyze the influence of the symmetry-breaking energy scale parameter 
$\eta$ on the shadow of the SV BH with PGM (see Figure \ref{figsh1}). The results reveal an asymmetric effect: while 
$\eta$ increases the shadow radius in OGM, it decreases the shadow radius in PGM. Additionally, for the same parameter values, the shadow size in OGM always remains larger than that in PGM.

 Now, we analysis the influence of the symmetry-breaking energy scale  parameter $\eta$ on the shadow of SV BH with PGM (see figure \ref{figsh1}). The  figure shows that  the influence of the parameter $\eta$ is  asymmetric: it enlarges the shadow radius in OGM but
diminishes it in PGM. Moreover, the shadow size in OGM is consistently larger than that
 in PGM for the same parameter values.

\begin{figure}[ht!]
    \centering
    \includegraphics[width=0.45\linewidth]{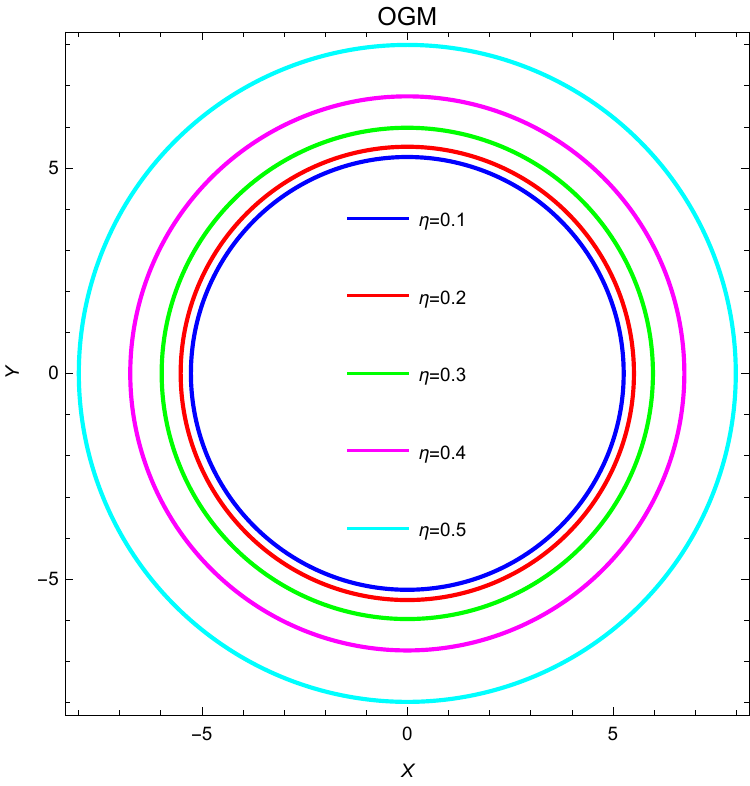}\quad\quad
    \includegraphics[width=0.45\linewidth]{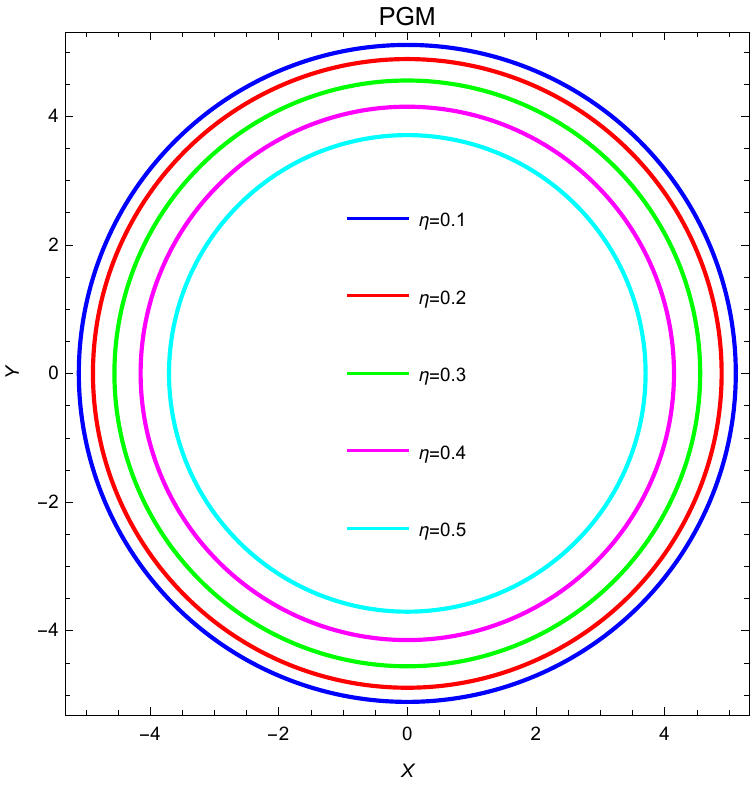}
    \caption{The geometrical shape of the shadow radius in terms of celestial coordinates for   several values of $\eta$ in the case of PGM (left), for OGM (right). {Here, we fix the observer at $r_o = 10^3 M, \quad \theta_o = \pi/2 $. }}
    \label{figsh1}
\end{figure}

\section{Conclusions and Summary} \label{sec:6}

In this study, we investigated the non-rotating SV BH solution in the presence of PGMs, exploring its thermodynamic properties and shadow. Our investigation rigorously characterized how the bounce  parameter $a$ and the PGM parameter $\eta$ affect thermodynamics and shadow, demonstrating considerable deviations from traditional GR predictions.

Our investigation commenced with the construction of a fundamental mathematical framework for SV BHs within the context of PGMs, leading to the derivation of the line element in Eq. (\ref{aa5}), where the metric function is defined by Eq. (\ref{aa6}). Analysis of the Ricci and Kretschmann scalars in Eqs. (\ref{scalar1})--(\ref{scalar2}) revealed that all curvature invariants remain finite everywhere, confirming the regularity of the BH solution. The event horizon radius demonstrated a significant dependence on both the bounce parameter 
$a$ and the PGM parameter 
$\eta$, with positive values of 
$\xi$ yielding larger horizons compared to negative values. Notably, we found that horizons in OGM consistently surpassed those in PGM under identical parameter conditions, as illustrated in Fig. \ref{figa1}.

The thermodynamic analysis of SV BH PGMs revealed profound insights into their stability and phase behavior. The Hawking temperature, derived in Eq. (\ref{Temperature}), exhibited an inverse dependence on the horizon radius, with temperature decreasing as the bounce parameter 
$a$ increased in PGM configurations (Fig. \ref{figa2}). Notably, the PGM case had a significantly stronger influence on Hawking temperature than the OGM case. The Bekenstein-Hawking entropy, given by Eq. (\ref{entr2}), remained independent of GM parameters but displayed modified scaling with horizon radius due to the bounce parameter. For global thermodynamic stability, we examined the Gibbs free energy and found that both the bounce parameter and the symmetry-breaking energy scale altered its behavior compared to the standard Schwarzschild BH. The PGM case induced a more pronounced modification than the OGM case. Furthermore, the Gibbs free energy remained positive, indicating that the SV BH—with or without GMs—is globally unstable. Regarding local stability, our analysis of specific heat uncovered a single divergence point (Figs. \ref{figa2c} and \ref{figa3c}), signaling a phase transition in the thermodynamic behavior of the SV BH with GMs. This distinction highlights a fundamental difference between PGM and OGM effects on BH physics.

Our analysis of the BH shadow yielded observationally relevant constraints on the theoretical parameters. We derived analytical expressions for both the photon sphere radius (Eq. (\ref{photons1})) and the shadow radius (Eq. (\ref{photons10})). The results reveal that the photon sphere radius 
$r_{ph}$
  depends on the bounce parameter 
$a$ and the energy scale of symmetry breaking 
$\eta$, whereas the critical impact parameter remains unaffected by 
$a$. Shadow key findings include:
\begin{itemize}
    \item Effect of the bounce parameter 
$a$: Increasing 
$a$ consistently reduces the photon sphere radius as seen in left panel of Figure \ref{figph1}.
\item Effect of the symmetry-breaking scale 
$\eta$: Its influence is asymmetric—enhancing the photon sphere radius in OGMs  but diminishing it in PGMs  as seen in right panel of Figure \ref{figph1}.
\item Shadow size dependence: The parameter 
$\eta$ decreases the shadow size in PGMs while increasing it in OGMs, as illustrated in Fig. \ref{figsh1}.
\end{itemize}

Our analysis revealed consistent patterns indicating that PGMs produce qualitatively distinct effects on BH physics compared to OGMs, despite their shared topological origins. Similarly, the bounce parameter $a$ systematically modifies all physical observables. The inclusion of bounce parameter and PGMs leads to profound changes in both the physical and geometric properties of the selected BH solution, distinguishing it markedly from the standard Schwarzschild BH. These differences provide new perspectives for understanding BHs in alternative gravitational theories.

This study paves the way for several promising research directions. Extending our analysis to rotating SV BHs with PGMs could yield insights into frame-dragging effects and ergosphere characteristics in these modified spacetimes \cite{isc1,isc2,isc3}. Additionally, investigating the interactions of SV BHs with PGM surrounding by dark matter may further elucidate their astrophysical relevance.

\section*{Data Availability Statement}

This manuscript has no associated data. [Authors’ comment: All data generated or analyzed during this study are included in the article.]

\section*{Code/Software Statement}

This manuscript has no associated code/software. [Authors’ comment: Code/Software sharing not applicable as no code/software was generated during in this current study.]

\section*{Conflict of Interests}

Author declare(s) no conflict of interest.

\end{document}